\documentclass[a4paper,11pt]{article}
\usepackage{jheppub} 
\usepackage{lineno,xcolor}

\DeclareMathOperator*{\diag}{diag}
\DeclareMathOperator*{\sgn}{sgn}

\renewcommand{\Im}{\operatorname{Im}}

\providecommand*{\iu}%
{\ensuremath{\mathrm{i}}}
\def\diff{\mathrm{d}}

\title{\boldmath Nonequilibrium steady states in driven holographic Weyl semi-metals}

\author[a,b]{Matteo Baggioli, }
\author[c]{Sebastian Grieninger, }
\author[d]{James Stokes}
\affiliation[a]{Wilczek Quantum Center, School of Physics and Astronomy, \\
Shanghai Jiao Tong University, Shanghai 200240, China}
\affiliation[b]{Shanghai Research Center for Quantum Sciences, Shanghai 201315, China}
\affiliation[c]{InQubator for Quantum Simulation (IQuS), Department of Physics, \\
University of Washington, Seattle, WA 98195, USA}
\affiliation[d]{Department of Mathematics, University of Michigan, Ann Arbor, MI 48109, USA}

\emailAdd{b.matteo@sjtu.edu.cn}
\emailAdd{segrie@uw.edu}
\emailAdd{stokesjd@umich.edu}

\abstract{Three-dimensional Weyl materials provide a controlled setting for exploring Floquet dynamics in open quantum systems, including nonequilibrium steady states (NESS). Motivated by the desire for a strongly-coupled description, we employ holography to analyze the formation and stability of a NESS in a Weyl semi-metal induced by an external circularly polarized electric field. A time-periodic steady-state solution is constructed and its stability is determined from the spectrum of out-of-equilibrium quasinormal modes (Floquet exponents). A stable region in the drive parameter space is identified; beyond a critical curve, the Floquet exponents enter the upper half of the complex plane, leading to a superharmonic response. At sufficiently strong driving, chaotic time evolution emerges in the fully nonlinear initial-boundary value problem. The anomaly-induced response of the NESS to an external magnetic field is also computed, and the resulting behavior is related to the previously proposed chiral pumping effect.}

\preprint{IQuS@UW-21-120, NT@UW-26-4}
\begin{document}
\maketitle
\flushbottom

\section{Introduction}

Floquet engineering exploits the response of a quantum many-body system to periodic driving in order to realize nonequilibrium states that would be difficult or impossible to achieve in a static setting. In isolated quantum systems, periodic driving typically leads to unbounded heating or, in special cases, to a discrete-time-crystalline (DTC) phase. In the presence of dissipation, however, the system typically settles into a nonequilibrium steady state (NESS) \cite{sato2025floquet}. Describing such states requires extending conventional thermodynamic and statistical frameworks beyond equilibrium. 

In the context of three-dimensional Dirac/Weyl materials \cite{RevModPhys.90.015001}, optical driving provides an effective approach for controlling electronic structure, while dissipation is a natural consequence of thermal contact with the environment. Experimentally, Weyl semi-metals may be produced from Dirac semi-metals by applying circularly polarized laser fields. The Dirac semi-metal consists of two massless Weyl fermions with opposite chirality located at the same momentum in the Brillouin zone. By applying a polarized laser field, for simplicity modeled by a rotating electric field, the Weyl fermions are then located at different momenta, forming a NESS referred to as Floquet Weyl semi-metal \cite{NIELSEN1983389,2014EL....10517004W,2016PhRvB..94w5137Z,2017NatCo...813940H,PhysRevB.96.041126}. The linearly dispersing nature of the quasiparticles near the Dirac/Weyl cones and the resulting vanishing density of states can produce conditions in which Coulomb interactions are weakly screened and interactions remain long-ranged. The extent to which this creates strongly-coupled dynamics is unsettled and likely material-dependent. If the material is conducive to strong interactions, then they are expected to play an important role in the NESS since they provide a mechanism by which the energy injected by the drive can be redistributed and ultimately dissipated. This motivates the use of a holographic framework that can reliably capture strongly coupled dynamics far from equilibrium~\cite{Hartnoll2016apf,Zaanen:2015oix,Baggioli:2019rrs,Natsuume:2014sfa}. 

In this work, we use holography to investigate the NESS associated with a Floquet Weyl semi-metal in which incident circularly polarized light triggers a transition from a topologically trivial phase to a gapless phase \cite{ebihara2016chiral}. In contrast to Ref.~\cite{ebihara2016chiral}, which derives an effective static Floquet Hamiltonian via a high-frequency expansion, the holographic framework does not rely on a Floquet expansion and enables access to the full time-dependent dynamics at arbitrary driving frequency, including the NESS microstructure.

Working in the probe approximation of the bottom-up model proposed in \cite{landsteiner2016holographic}, the dynamical formation of the NESS is studied and its stability is analyzed. The holographic model predicts that a stable NESS develops for a wide range of parameters, and we characterize the corresponding region of parameter space by studying the spectrum of linear fluctuations. Outside the stable region, the NESS becomes unstable and develops superharmonic periodic oscillations. For sufficiently strong driving, the time development becomes chaotic. As an additional contribution, we explore the response of the NESS to an external magnetic field. The findings from holography are shown to be consistent with the expectation from free Dirac theory.

Weyl semi-metals have been realized within holography using both bottom-up \cite{landsteiner2016holographic} and top-down \cite{hashimoto2017holographic, kinoshita2018holographic, fadafan2021weyl} constructions. The most closely related work is Ref.~\cite{hashimoto2017holographic}, which investigates a Floquet Weyl semi-metal dual to D7-branes in $\textrm{AdS}_5 \times S^5$ and demonstrates the existence of a NESS\footnote{Holographic Weyl semi-metals have been investigated in various directions in, e.g., ~\cite{Ahn:2024ozz,Rai:2024bnr,Copetti:2016ewq,Landsteiner:2016stv,Ji:2021aan,Landsteiner:2014vua,Copetti:2019rfp,Liu:2018djq,Liu:2018spp,Ammon:2018wzb,baggioli2021detecting,baggioli2018conjecture,baggioli2023entanglement}.}. Their analysis, however, does not examine how this NESS forms out of equilibrium or whether it is dynamically stable. Holographic realizations of NESS have also been investigated in several directions, e.g., \cite{Chang:2013gba,Amado:2015uza,Sonner:2017jcf,Ecker:2021ukv,Bachas:2021tnp,Kundu:2013eba,Brattan:2024dfv}.

\section{Chiral pumping effect}\label{sec:CPE}
Quantum anomalies generate contributions to transport in relativistic quantum matter \cite{landsteiner2016notes}.
 In a $U(1)_V \times U(1)_A$ gauge theory in $3+1$ dimensions, Lorentz covariance and $U(1)_V$ gauge invariance constrain the form of the anomaly-induced vector current to leading order in the axial background $A$ as follows,
\begin{equation}\label{e:tensor_structure}
    J_V^a \propto \epsilon^{abcd} A_b H_{cd},
\end{equation}
where $H_{ab}=\partial_{a} V_b - \partial_b V_a$ is the vector field strength and the coefficient is fixed by the anomaly. The chiral magnetic effect (CME) \cite{kharzeev2008effects,fukushima2008chiral} follows by specializing to an axial background of the form $A = \mu_A \diff x^0$ and switching on a magnetic field $\vec{B}$, yielding a vector current along the field,
\begin{equation}
    \vec{J}_V \propto \mu_A \vec{B} \quad \textrm{(CME)}.
\end{equation}
The same tensor structure \eqref{e:tensor_structure}
suggests a response of the vector charge density $\rho$ when the axial background is spacelike $A = \vec{A}_5 \cdot \diff \vec{x}$, which has been referred to as the chiral pumping effect \cite{ebihara2016chiral},
\begin{align}
    \rho & \propto \epsilon^{0ijk}A_i H_{jk}  \propto \vec{A}_5 \cdot \vec{B} \quad \textrm{(CPE)}.
\end{align}
The terminology emphasizes that, since the vector charge is conserved, the relation above must be interpreted in an open quantum system setting in which charge is pumped into the sample from an external reservoir. Moreover, in the presence of a magnetic field, the dynamics can be understood via an effective $(1+1)$-dimensional picture associated with the lowest-Landau-level, which becomes exact in the limit of strong magnetic field. 
The resulting chiral one-dimensional theory predicts that the pumped vector density is accompanied by an axial current aligned with the magnetic field. In contrast to the charge density, however, the axial current is not generically anomaly-protected and can receive material-dependent contributions from higher Landau levels. These expectations can be confirmed either by an explicit field-theoretic calculation in Dirac field theory \cite{ebihara2016chiral} or by a complementary first-quantized analysis, which we provide in appendix \ref{app:CPE}.

In the context of Weyl semi-metals, an axial gauge field provides an effective field theory parametrization of the Weyl-node separation in energy-momentum space \cite{ebihara2016chiral}. A spatial $\vec{A}_5$ shifts the nodes in opposite directions in momentum space, breaking time-reversal symmetry (TRS). Periodic optical driving provides a controlled way to generate such a background. In particular, when TRS is broken by circularly polarized light with angular frequency $\omega_\textrm{D}$, the driven dynamics are conveniently described in a Floquet framework, and on observation times that are large compared to the period $2\pi/\omega_{\mathrm D}$ one obtains an effective (cycle-averaged) Hamiltonian containing an emergent axial field $\vec{A}_5$ oriented orthogonally to the polarization plane.

\section{Holographic model}\label{sec:model}
A Weyl semi-metal is characterized by a $U(1)_V \times U(1)_A$ symmetry corresponding to a conserved vector and anomalous axial current. 
A minimal holographic quantum field theory (QFT) that realizes this symmetry \cite{landsteiner2016holographic} can be obtained by considering a $(4+1)$-dimensional asymptotically anti-de Sitter spacetime with a $(3+1)$-dimensional conformal boundary on which the holographic QFT lives. The bulk spacetime contains a pair of abelian gauge fields $(V_\mu,A_\mu)$ coupled via a Chern-Simons interaction which encodes the $U(1)_V \times U(1)_A$ anomaly structure, as well as a scalar field charged under $U(1)_A$. In the probe limit, where the metric is treated as a nondynamical background, the action associated with a spacetime volume $\mathcal{V}$ is given by
\begin{equation}
    S = S_\textrm{MCS} + S_\textrm{KG},
\end{equation}
with
\begin{align}
    S_\textrm{MCS} 
    & = \int_\mathcal{V} \diff^5 x \sqrt{-g}\left[- \frac{1}{4}F^{\mu\nu}F_{\mu\nu} - \frac{1}{4} H^{\mu\nu}H_{\mu\nu} + \frac{\alpha}{3} \bar\epsilon^{\mu\nu\rho\lambda\sigma}A_\mu(F_{\nu\rho}F_{\lambda\sigma}+3H_{\nu\rho}H_{\lambda\sigma})\right], \\
    S_\textrm{KG}
    & = - \int_\mathcal{V} \diff^5 x \sqrt{-g} \left[ g^{\mu\nu} (D_\mu \phi)^\ast D_\nu \phi + m^2 \phi^\ast \phi \right],
\end{align}
and where $F_{\mu\nu} = \partial_\mu A_\nu - \partial_\nu A_\mu$, $H_{\mu\nu} = \partial_\mu V_\nu - \partial_\nu V_\mu$, $D_\mu = \partial_\mu - \iu q_5 A_\mu$ and the  Levi-Civita tensor density is defined as $\bar\epsilon^{\mu\nu\rho\lambda\sigma} = \epsilon^{\mu\nu\rho\lambda\sigma} / \sqrt{-g}$ where $\epsilon^{\mu\nu\rho\lambda\sigma}$ is the standard permutation symbol. 

The corresponding equations of motion are
\begin{align}
    0 & = g^{\mu\nu} \big(\nabla_\mu - \iu q_5 A_\mu \big) D_\nu\phi - m^2\phi, \\
    0 & =\frac{1}{\sqrt{-g}}\partial_\nu\big(\sqrt{-g} F^{\nu\mu}\big) + \alpha \epsilon^{\mu\nu\rho\lambda\sigma}(F_{\nu\rho}F_{\lambda\sigma}+H_{\nu\rho}H_{\lambda\sigma}) + \iu q_5 g^{\mu\nu}\big[\phi (D_\nu\phi)^\ast-\phi^\ast D_\nu\phi\big], \\
    0 & = \frac{1}{\sqrt{-g}}\partial_\nu\big(\sqrt{-g} H^{\nu\mu}\big) + 2\alpha \epsilon^{\mu\nu\rho\lambda\sigma}F_{\nu\rho}H_{\lambda\sigma}.
\end{align}

In order to prevent uncontrolled heating due to periodic driving, we consider states of the QFT at fixed temperature, which corresponds to choosing the bulk spacetime to be an anti-de Sitter black hole. Energy injected by the drive is absorbed by the horizon, which, in the probe limit, does not backreact and therefore does not change the  temperature.

It will prove convenient to work in the ingoing Eddington-Finkelstein coordinate system
\begin{equation}
    x^\mu=(t,z,x^1,x^2,x^3) = (t,z,\vec{x}) \in \mathbb{R}\times (0,1] \times \mathbb{R}^3,
\end{equation}
in which the line element is given by
\begin{align}
    \diff s^2 & = \frac{1}{z^2}\big({-}f(z)\diff t^2 - 2 \diff t \diff z + \diff \vec x^2\big), \\
    f(z) & = 1-z^4,
\end{align}
where we have fixed the temperature by demanding that the horizon is located at $z=1$.

In order to fix the background metric for the QFT, we choose the spacetime volume $\mathcal{V}$ to be the region $z \geq \epsilon$ where $0<\epsilon \ll 1$. It follows that the induced metric $\gamma_{ab}$ on the boundary spacetime $\partial \mathcal{V}$ is given by
\begin{equation}
    \gamma_{ab} = \frac{1}{\epsilon^2}\diag\big({-f(\epsilon),1,1,1}\big),
\end{equation}
which approaches $\gamma_{ab} \sim \frac{1}{\epsilon^2}\eta_{ab}$, thereby fixing the representative of the conformal class to the flat Minkowski metric. In order to obtain a finite $\epsilon \to 0$ limit, holographic renormalization dictates that the action must be supplemented by the following counterterms (assuming $m^2=-3$),
\begin{align}
    S_{\rm ct} & = \int_{\partial\mathcal{V}} \diff^4 x \sqrt{-\gamma} \left\{ 
    \left[\gamma^{ab}\phi^\ast D_a D_b \phi-\frac{1}{4} F^2-\frac{1}{4}H^2\right] \log\epsilon
    -
    \phi^\ast \phi
    \right\}.
\end{align}

\subsection{Background field configuration}
In terms of the Eddington-Finkelstein coordinates above, we impose the following gauge condition for the vector and axial gauge fields,
\begin{equation}\label{e:radialgauge}
    A_z=V_z=0.
\end{equation}

In Ref. \cite{ebihara2016chiral}, the chiral pumping effect was induced using a combination of a rotating electric field in the $x^1$-$x^2$ plane combined with a uniform magnetic field in the $x^3$ direction. The corresponding background ansatz for the 5D fields is determined to be
\begin{align}
    \phi & = \phi(z,t) \in \mathbb{R}, \\
    A_\mu & = \big(0,0,0,0, A_3(z,t) \big), \\
    V_\mu & = 
    \big( V_0(z,t),0, V_1(z,t), V_2(z,t),0 \big)
    +
    \big(0,0,0,Bx^1,0\big).
\end{align}
The background equations of motion are then
\begin{align}
    z^2 f \phi'' + 3z\dot{\phi} + \left( z^2 f'  - 3z f \right) \phi' - 2 z^2\dot{\phi}' - \left( m^2 + q_5^2 z^2  A_3^2 \right) \phi & = 0, \label{e:bg1} \\
    z^2 f  A_3'' + z \dot{ A}_3 + \left( z^2 f'  - z f \right)  A_3' - 2 z^2 \dot{ A}_3' - 2 q_5^2  A_3  \phi^2 + 8 z^3\alpha\left(  V_1' \dot{ V}_2 -  V_2' \dot{ V}_1-B V_0'\right) & = 0, \label{e:bg2} \\
    z f  V_1'' + \left( z f' - f \right)   V_1' + \dot{ V}_1 -2 z \dot{ V}_1' + 8z^2 \alpha \left(  V_2'\dot{ A}_3-  A_3'\dot{ V}_2\right) & = 0, \label{e:bg3} \\
    z f  V_2'' + \left( z f' - f \right)  V_2' + \dot{ V}_2 -2 z \dot{ V}_2' - 8z^2 \alpha \left(  V_1' \dot{ A}_3-  A_3'\dot{ V}_1\right) & = 0, \label{e:bg4} \\
    z V_0'' -  V_0' - 8z^2 \alpha B  A_3' & = 0, \label{e:bg5} \\
    \dot{ V}'_0 - 8 z \alpha B \dot{ A}_3 & = 0. \label{e:bg6}
\end{align}
It should be emphasized that a nontrivial solution supporting a nonzero magnetic field necessitates the introduction of the potential ${V}_0$. The ultraviolet (UV) boundary conditions at $z=0$ are given by
\begin{align}
    \phi'(t,0)
    & = M, & A_3(t,0) & = b, & V_1(t,0) & = S_1(t), \\
    V_2(t,0) & = S_2(t), & V_0(t,0) & = S_0(t),
\end{align}
and the infrared (IR) boundary conditions are determined by regularity at the horizon. In particular, they follow from evaluating bulk equations of motion at $z=1$.

The general solution of the background equations subject to the UV boundary conditions is
\begin{align}
     \phi(z,t)
    & =
    M z
    +
    \left[\varphi(t)+\frac{1}{2}q_5^2 b^2 M \log z\right]z^3 + \cdots, \\
     A_3(z,t)
    & = b +
    \left[a(t)+q_5^2 b M^2\log z\right]z^2 + \cdots, \\
     V_1(z,t)
    & =
    S_1(t) + \dot{S}_1(t)z + \left[v_1(t)+ \frac{1}{2}\ddot{S}_1(t)\log z\right]z^2 + \cdots, \\
     V_2(z,t)
    & =
    S_2(t) + \dot{S}_2(t)z + \left[v_2(t)+ \frac{1}{2}\ddot{S}_2(t)\log z\right]z^2 + \cdots, \\
     V_0(z,t)
    & =
    S_0(t)
    - \frac{1}{2} \rho  z^2
    + 8\alpha B\int_0^z \diff y \, y \,   A_3(y,t). \label{e:V0}
\end{align}
Note that in the radial gauge \eqref{e:radialgauge}, there is a residual gauge symmetry $V_\mu \to V_\mu + \partial_\mu \lambda$ with gauge parameter of the form $\lambda=\lambda(t,\vec{x})$. Restricting to gauge parameters that preserve the $\vec{x}$-independent ansatz subspace implies $\lambda = \lambda(t)$ and thus $V_{1,2}$ are gauge-invariant, while $V_0$ is gauge-variant. Thus, rather than fixing the gauge-variant boundary condition $S_0$, it is much more convenient to fix the gauge-invariant potential difference (taken between the boundary and the horizon for convenience),
\begin{align}\label{e:potential_difference}
    \Delta V_0(t) 
    & := V_0(z=0,t) - V_0(z=1,t), \\
    & = \frac{1}{2}\rho  - 8 \alpha B \int_0^1 \diff z \, z \, A_3(z,t).
\end{align}
The general solution is then parametrized in terms of the following UV data, and the remaining sub-leading data are determined by imposing regularity at the horizon:
\begin{equation}
  \text{Input (UV) data:}\quad   \left\{
    \begin{aligned}
        &M \in \mathbb{R}\\
        &b \in \mathbb{R} \\
        &S_1=S_1(t) \\
        &S_2=S_2(t) \\
        &\Delta V_0 = \Delta V_0(t)
    \end{aligned}
    \right.;
    \qquad
    \text{Output (sub-leading) data:}\quad 
    \left\{
    \begin{aligned}
        &\varphi=\varphi(t) \\
        &a=a(t) \\
        &v_1=v_1(t) \\
        &v_2=v_2(t) \\
        &\rho \in \mathbb{R}
    \end{aligned}
    \right.
\end{equation}

A standard calculation in holographic renormalization (reviewed in appendix \ref{app:renormalization}) relates the above data to the one-point functions of the dual currents,
\begin{align}
    \langle J_V^0 \rangle_\textrm{ren} & = \rho , \\
    \langle J_V^1 \rangle_\textrm{ren} & = 2v_1(t) - 8b\alpha \dot{S}_2(t) - \frac{1}{2}\ddot{S}_1(t), \\
    \langle J_V^2 \rangle_\textrm{ren} & = 2v_2(t) + 8b\alpha \dot{S}_1(t) - \frac{1}{2}\ddot{S}_2(t), \\
    \langle J_A^3 \rangle_\textrm{ren} & = 2 a(t) + b q_5^2 M^2.
\end{align}

\subsection{Anomalous Hall effect}\label{sec:ahe}
In this section we revisit the anomalous Hall effect as a concrete example of holographic renormalization in Eddington-Finkelstein coordinates. In particular, we show how the renormalized boundary current is obtained unambiguously in the presence of the Chern-Simons coupling and how this yields the standard relation between the Hall response and the horizon value of the axial gauge field.

In order to reproduce the anomalous Hall effect, consider $B=0$ and assume that $(\phi,A_3)$ are independent of time. Then the equations of motion become
\begin{align}
    z^2 f \phi'' + \left( z^2 f'  - 3z f \right) \phi' - \left( m^2 + q_5^2 z^2  A_3^2 \right) \phi & = 0, \\
    z^2 f  A_3'' + \left( z^2 f'  - z f \right)  A_3' - 2 q_5^2  A_3  \phi^2 & = 0, \\
    z f  V_1'' + \left( z f' - f \right)   V_1' + \dot{V}_1 -2 z \dot{V}_1' - 8z^2 \alpha A_3'\dot{V}_2 & = 0, \\
    z f  V_2'' + \left( z f' - f \right)  V_2' + \dot{V}_2 -2 z \dot{V}_2' + 8z^2 \alpha A_3'\dot{V}_1 & = 0.
\end{align}
Now consider the response of the current to a harmonic electric field.
It is convenient to define $V_\pm = V_1 \pm \iu V_2$ and to implement the source using a harmonic ansatz $V_\pm(z,t)= e^{-\iu \omega t} v_\pm(z)$. Then we find
\begin{align}
    z^2\left(\frac{f}{z}v_\pm'\right)' -\iu \omega (v_\pm -2 z v_\pm') \pm 8z^2 \omega \alpha A_3'v_\pm & = 0,
\end{align}
Expanding in the angular frequency $v_\pm = v_\pm^{(0)} + \omega v_\pm^{(1)} + \cdots$ we obtain
\begin{equation}
    v_\pm^{(0)} = c_\pm \in \mathbb{C},
\end{equation}
and
\begin{equation}
    \left(\frac{f}{z}(v^{(1)}_\pm)' \pm 8 c_\pm \alpha A_3\right)' - \frac{\iu c_\pm}{z^2} = 0.
\end{equation}
Integrating this expression from an arbitrary bulk point to horizon yields a first-order differential equation for $v^{(1)}_\pm$, which is solved by
\begin{align}
    v_\pm^{(1)}(z)-v_\pm^{(1)}(0)
    & =
    c_\pm \int_0^z \frac{\diff y}{f(y)}
    \left[
    \iu(y-1)\pm 8\alpha y\big(A_3(1)-A_3(y)\big)
    \right], \\
    & = c_\pm \left[-\iu z + \frac{1}{2}\big(\iu \pm 8\alpha(A_3(1)-b)\big)z^2 + \cdots\right].
\end{align}
Choosing the integration constant to be $v_\pm^{(1)}(0)=0$ ensures that the source is determined by $v_\pm^{(0)}$. In particular, $S_{j}(t)= c_{j}e^{-\iu \omega t}$ where $c_\pm = c_1 + \iu c_2$ and $c_{1,2}\in\mathbb{R}$. The boundary electric field is likewise given by $E_{j}(t)=-S_{j}'(t) = \iu \omega c_{j}e^{-\iu \omega t}$. Choosing the electric field to be aligned in the $x^1$-direction ($c_2=0$ and $c_1 \in \mathbb{R}$) we obtain the following conductivities
\begin{align}
    \lim_{\omega \to 0} \frac{\langle J_V^1 \rangle_\textrm{ren}}{E_1} & = 1, \\
    \lim_{\omega \to 0} \frac{\langle J_V^2 \rangle_\textrm{ren}}{E_1} & = - 8 \alpha A_3(1).
\end{align}
Comparing with the prediction of Dirac theory \cite{landsteiner2016holographic}, one concludes that $A_3(1) = b_\textrm{eff}$ where $b_\textrm{eff}$ is the effective chiral shift. 

\section{Numerical experiments}\label{sec:experiments}
In our numerical simulations we consider the following default parameters corresponding to the gapped phase ($M=\alpha=q_5=1$, $b=0$). All dimensionful parameters are understood to be normalized with respect to $M$.

\subsection{Periodic driving by electric field}\label{sec:NESS}
In order to realize a nonequilibrium steady state (NESS), we consider the fully time-dependent problem with $B=0$. The relevant equations of motion are now
\begin{align}
    z^2 f \phi'' + 3z\dot{\phi} + \left( z^2 f'  - 3z f \right) \phi' - 2 z^2\dot{\phi}' - \left( m^2 + q_5^2 z^2  A_3^2 \right) \phi & = 0, \\
    z^2 f  A_3'' + z \dot{ A}_3 + \left( z^2 f'  - z f \right)  A_3' - 2 z^2 \dot{ A}_3' - 2 q_5^2  A_3  \phi^2 + 8 z^3\alpha\left(  V_1' \dot{ V}_2 -  V_2' \dot{ V}_1\right) & = 0, \\
    z f  V_1'' + \left( z f' - f \right)   V_1' + \dot{ V}_1 -2 z \dot{ V}_1' + 8z^2 \alpha \left(  V_2'\dot{ A}_3-  A_3'\dot{ V}_2\right) & = 0, \\
    z f  V_2'' + \left( z f' - f \right)  V_2' + \dot{ V}_2 -2 z \dot{ V}_2' - 8z^2 \alpha \left(  V_1' \dot{ A}_3-  A_3'\dot{ V}_1\right) & = 0,
\end{align}
Consider an electric field rotating at angular frequency $\omega_\textrm{D}$. The existence of the NESS was predicted in free Dirac theory \cite{ebihara2016chiral} for large $\omega_\textrm{D} \gg 1$ on the basis of Floquet perturbation theory. The holographic QFT discussed here, in contrast, is strongly interacting and our considerations are fully non-perturbative in $\omega_\textrm{D}$. In particular, this enables exploration of the non-perturbative stability of the NESS. 
The boundary conditions on the vector gauge field are
\begin{align}
    S_1(t) & = P \cos(\omega_\textrm{D} t),  & S_2(t) & = P \sin(\omega_\textrm{D} t).
\end{align}

\begin{figure}
\centering
\includegraphics[width=0.4\textwidth]{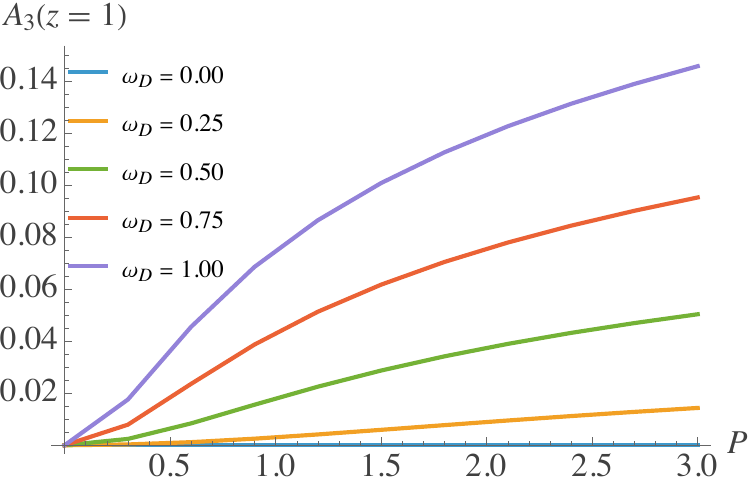}
\includegraphics[width=0.4\textwidth]{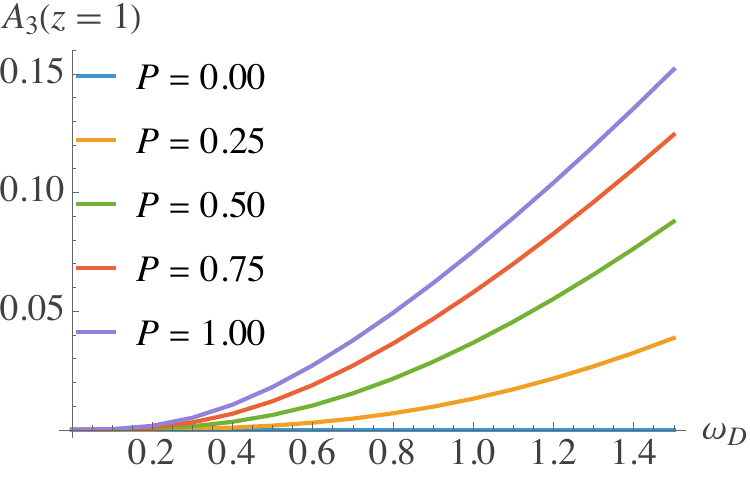}
\includegraphics[width=0.4\textwidth]{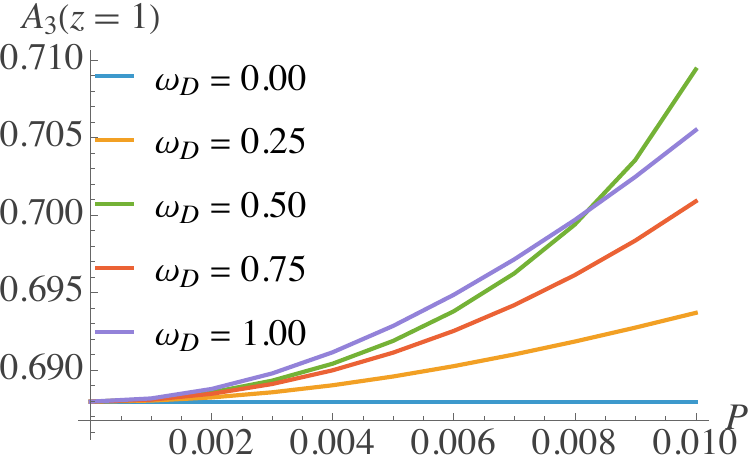}
\includegraphics[width=0.4\textwidth]{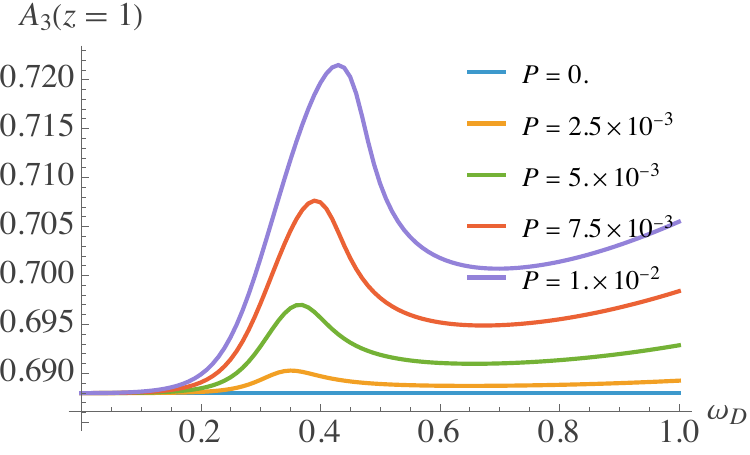}
\caption{The response of $A_3(z=1)$ to the driving parameters $(P,\omega_\textrm{D})$.
Top row ($\alpha=1$, $b=0$): Monotonic behavior. Bottom row ($\alpha=10$, $b=1$) Non-monotonic behavior with level crossings.
In contrast to free fermions \cite{ebihara2016chiral}, we do not make assumptions on the largeness of $\omega_\textrm{D}$.\label{fig:staticresponse}}
\end{figure}

In order to find the NESS predicted by \cite{ebihara2016chiral}, we utilize the following ansatz,
\begin{equation}\label{e:ansatz}
    \left\{
    \begin{aligned}
    \phi(z,t) & = \phi(z), \\
    A_3(z,t) & = A_3(z), \\
    V_\pm(z,t) & = g_\pm(z) e^{\pm \iu \omega_\textrm{D} t}
    \end{aligned}
    \right.
\end{equation}
where $g_\pm(z) = g_1(z) \pm \iu g_2(z)$ and $g_{1,2}(z) \in \mathbb{R}$.
It should be emphasized that this regime is not accessible to the free fermion calculation of \cite{ebihara2016chiral}, which assumes a large-$\omega_\textrm{D}$ expansion. 

The response of the horizon-value of the axial gauge field $A_3(z=1)$ to the driving parameters is illustrated in Fig.~\ref{fig:staticresponse}. Fixing the UV value $b = A_3(z=0) = 0$ and setting the Chern-Simons coupling to $\alpha=1$, we find that $A_3(z=1)$ varies monotonically with both the amplitude and the angular frequency. Interestingly, this monotonic behavior is not generic: for sufficiently large Chern-Simons coupling and nonzero $b$, the monotonic dependence breaks down and a peak develops as a function of the angular frequency.

In order to understand the stability of the NESS, which was not addressed in \cite{ebihara2016chiral}, we study linear fluctuations about the time-periodic background \eqref{e:ansatz}. Since the time-periodic NESS does not admit a timelike Killing vector, the standard notion of quasinormal modes (QNMs) is not applicable. Nevertheless, one can consider an out-of-equilibrium generalization of QNMs, which corresponds to generalizing the ansatz \eqref{e:ansatz} via the following replacements,
\begin{align}
    \phi(z) & \to \phi(z) + e^{-\iu \omega t} \delta \phi(z), \\
    A_3(z) & \to A_3(z) + e^{-\iu \omega t} \delta A_3(z), \\
    g_\pm(z) & \to g_\pm (z) + e^{-\iu \omega t} \delta g_\pm(z),
\end{align}
where $\omega \in \mathbb{C}$ are called Floquet exponents. Despite the time-dependence of the background, its discrete time-translation invariance implies that the periodic time dependence factors out, so that $\omega$ can be computed using the standard QNM technology.

\begin{figure}[h]
\centering
\includegraphics[width=0.4\textwidth]{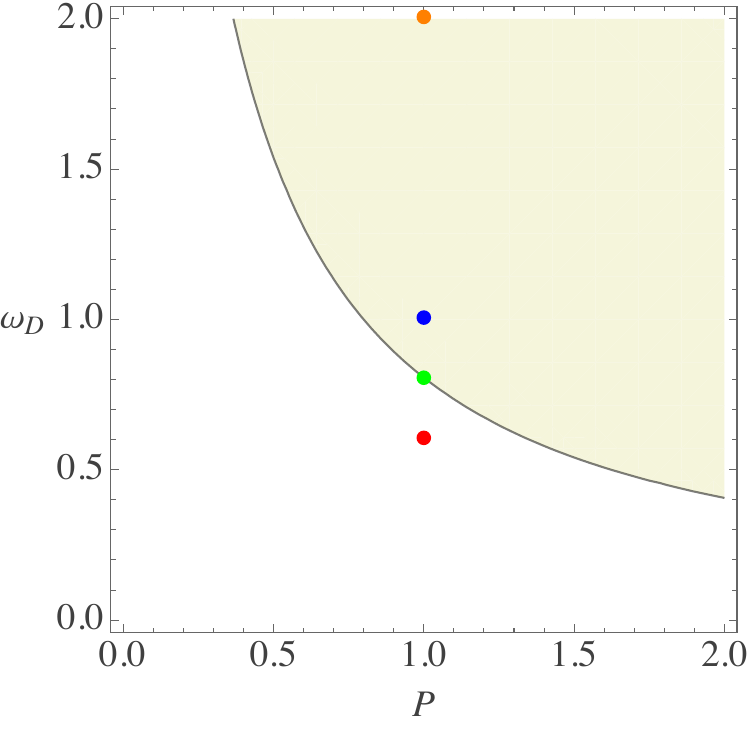}
\includegraphics[width=0.4\textwidth]{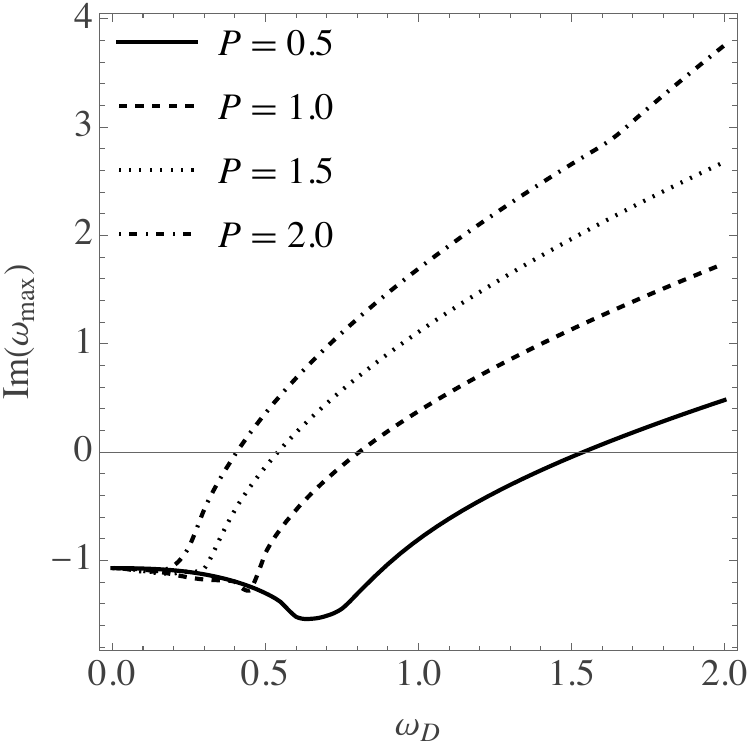}
\caption{Left: Phase diagram showing the stability/instability transition of the NESS with $B=0$. The beige region indicates that the dominant Floquet exponent $\omega_{\max}$ has entered the upper half of the complex plane. Right: Cross-sections showing $\Im(\omega_{\max})$ as a function of driving frequency $\omega_\textrm{D}$ for fixed amplitude $P$.
\label{fig:phase}}
\end{figure}

Analyzing the dominant Floquet exponent $\omega_{\max}$ as a function of the drive parameters one identifies a critical curve in $(P,\omega_\textrm{D})$-space (see Fig.~\ref{fig:phase}). The critical curve separates a stable region at weak driving characterized by $\Im(\omega_{\max})<0$ from an unstable region at strong driving in which $\omega_{\max}$ enters the upper half of the complex plane. In addition, we visualize the passage of $\omega_{\max}$ from the lower to the upper half of the complex plane as a function of driving frequency $\omega_\textrm{D}$, holding the amplitude $P$ constant.

\begin{figure}[h]
\centering
\includegraphics[width=\textwidth]{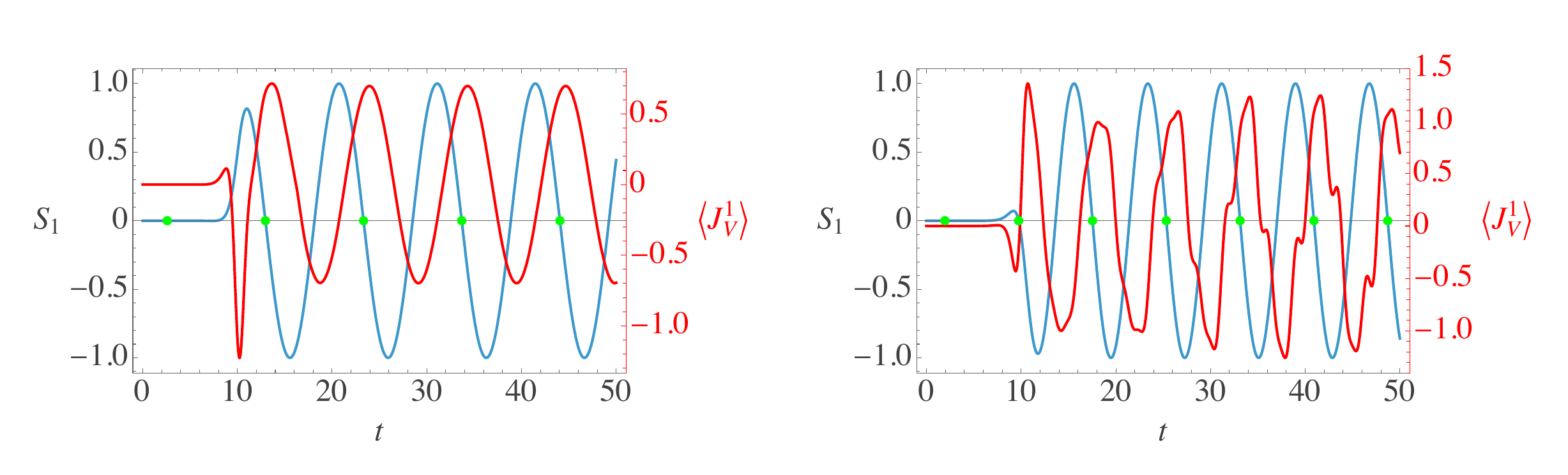}
\includegraphics[width=\textwidth]{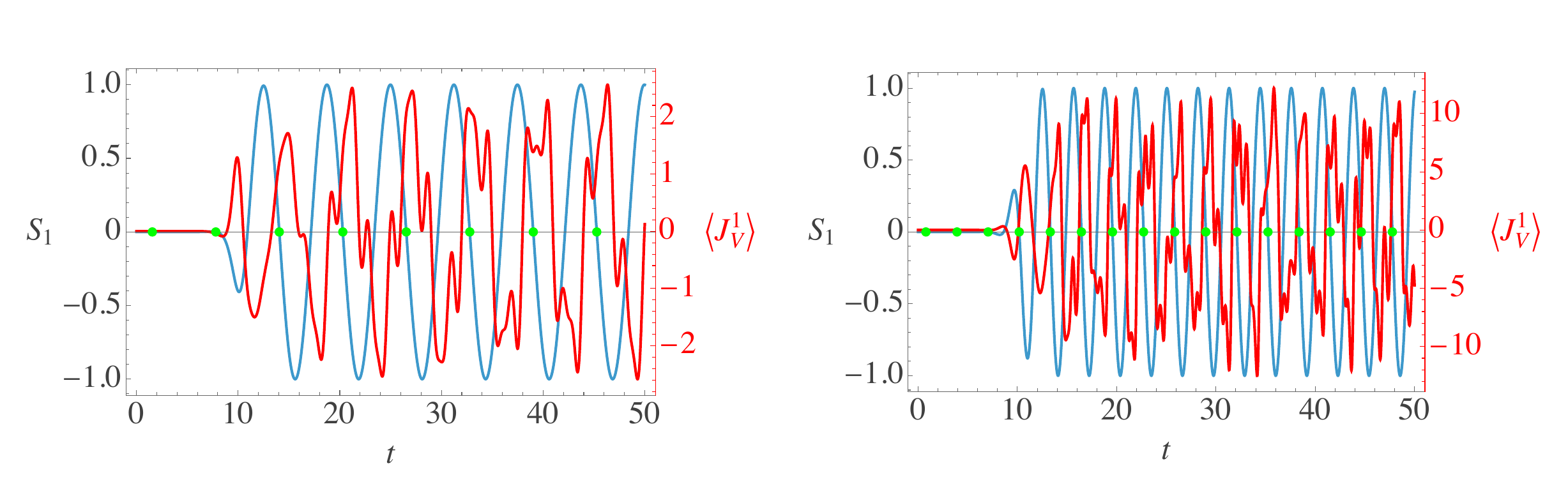}
\caption{
From left to right: Time evolution of the vector response corresponding to the points labeled from bottom to top in Fig. \ref{fig:phase}. The source is turned on at time $T=10$. The source $S_1$ is labeled in blue and the vector response $\langle J_V^1\rangle_\textrm{ren}$ is labeled in red. The vector response is evidently commensurate with the source. \label{fig:IVP_Vector}}
\end{figure}

\begin{figure}[h]
\centering
\includegraphics[width=\textwidth]{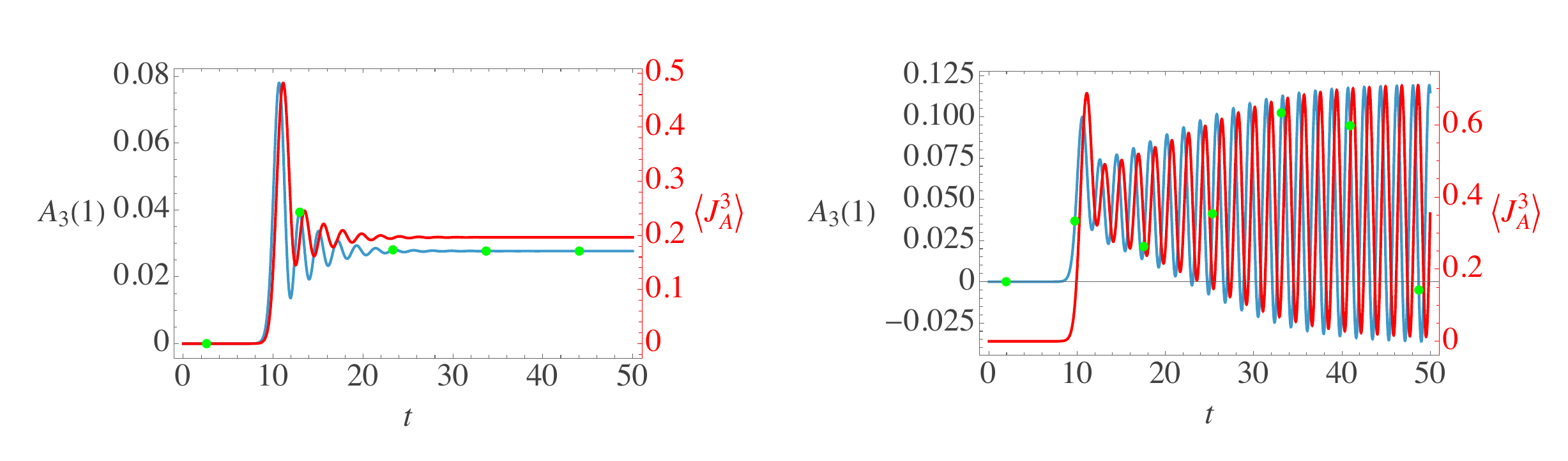}
\includegraphics[width=\textwidth]{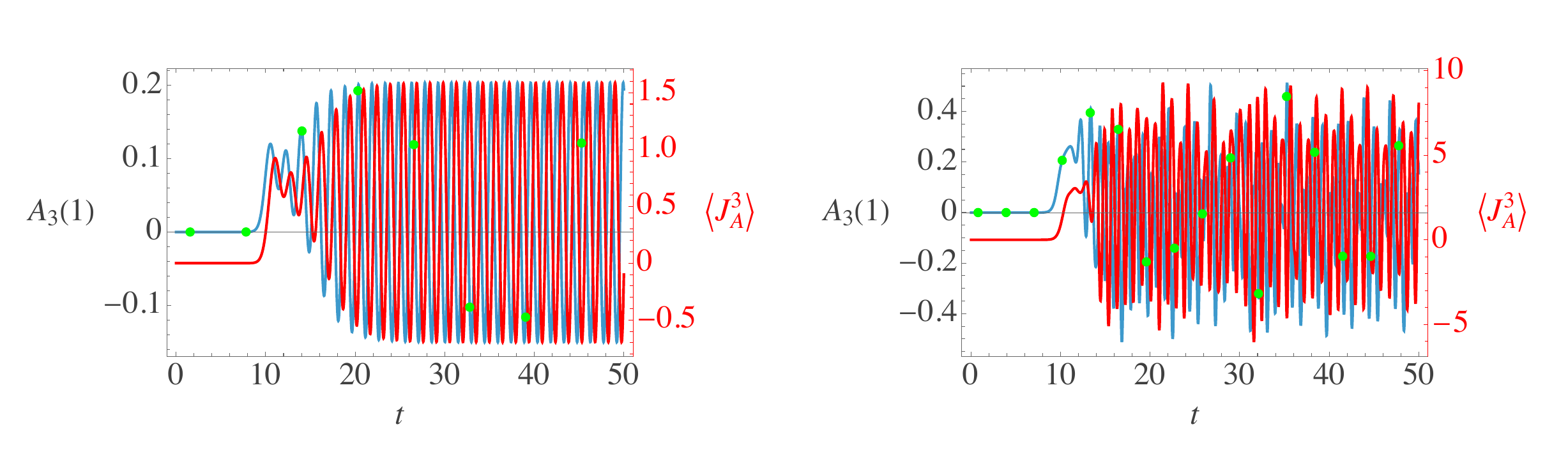}
\caption{
From left to right: Time evolution of the axial response corresponding to the points labeled from bottom to top in Fig. \ref{fig:phase}. The source is turned on at time $T=10$. The horizon-value of the gauge field  $A_3(z=1)$ is labeled in blue and the axial current $\langle J_A^3\rangle_\textrm{ren}$ is labeled in red. Despite not corresponding to a quantum observable, $A_3(z=1)$ provides an effective probe of the nonequilibrium stability of the NESS. The overlaid green dots indicate the temporal position of the nodes of the vector source $S_1$. Unlike the vector response, the axial response is clearly superharmonic.\label{fig:IVP_Axial}}
\end{figure}

In order to gain additional insight about the nature of the NESS and its destabilization, we solve the initial-boundary value problem with the static initial condition and subjected to a periodic drive that turns on at a critical time $T$ (chosen to be small compared to the total simulation time)\footnote{Additional details about the initial-boundary value problem formulation are presented in appendix \ref{app:IBVP}.}. In particular, the sources are given by
\begin{align}
    S_1(t) & = P\left[\frac{1+\tanh(t-T)}{2}\right]\cos\omega_\textrm{D} t, &
    S_2(t) & = P\left[\frac{1+\tanh(t-T)}{2}\right]\sin\omega_\textrm{D} t.
\end{align}
The evolution of the axial and vector current responses obtained by solving the initial-boundary value problem are presented in Figs.~\ref{fig:IVP_Vector} and \ref{fig:IVP_Axial}. The time evolution reveals that in the vicinity of the critical curve, the oscillations average out to give an effectively nonzero Weyl distance. For sufficiently strong driving, the nature of the time evolution appears to be chaotic, which calls into question the validity of the probe approximation deep in the unstable regime. A notable feature of the response is that, while the vector currents are commensurate with the drive, the axial response is distinctly superharmonic. Appendix \ref{app:superharmonic} provides a possible field-theoretic explanation of this phenomenon, although a general argument able to predict the precise superharmonic frequencies in the strongly coupled theory is not available.

\subsection{Response to magnetic field}
Now we investigate the anomaly-induced response of the Floquet NESS to a static external magnetic field $\vec{B} = B \hat{e}_3$. Recall that the chiral pumping effect predicts that the drive-generated axial background produces, in the presence of $\vec{B}$, a vector charge density and an axial current aligned with $\vec{B}$. In the holographic model, these observables are detected by the following one-point functions, respectively,
\begin{equation}
    \langle J_V^0 \rangle_\textrm{ren} = \rho , \qquad
    \langle J_A^3 \rangle_\textrm{ren} = 2 a(t) + b q_5^2 M^2.
\end{equation}
In the canonical parameter choice with $b=0$, the latter equation reduces to $\langle J_A^3 \rangle_\textrm{ren} = 2 a(t)$.

Integrating the background equation \eqref{e:bg6} with respect to time, one can eliminate $V_0$ from \eqref{e:bg1}--\eqref{e:bg5}, thereby obtaining a decoupled system for the fields $(\phi,A_3,V_1,V_2)$, which is parametrized by $(\rho ,B) \in \mathbb{R}^2$. This yields the following integro-differential system,
\begin{align}
    0 & = z^2 f \phi'' + 3z\dot{\phi} + \left( z^2 f'  - 3z f \right) \phi' - 2 z^2\dot{\phi}' - \left( m^2 + q_5^2 z^2  A_3^2 \right) \phi, \label{e:cpe1} \\
    0 & = z^2 f  A_3'' + z \dot{ A}_3 + \left( z^2 f'  - z f \right)  A_3' - 2 z^2 \dot{ A}_3' - 2 q_5^2  A_3  \phi^2 + {} \notag\\
    & \quad +8 z^3\alpha\left(  V_1' \dot{ V}_2 -  V_2' \dot{ V}_1-B (8 \alpha B z  A_3 - \rho  z)\right), \label{e:cpe2} \\
    0 & =z f  V_1'' + \left( z f' - f \right)   V_1' + \dot{ V}_1 -2 z \dot{ V}_1' + 8z^2 \alpha \left(  V_2'\dot{ A}_3-  A_3'\dot{ V}_2\right) , \label{e:cpe3} \\
    0 & = z f  V_2'' + \left( z f' - f \right)  V_2' + \dot{ V}_2 -2 z \dot{ V}_2' - 8z^2 \alpha \left(  V_1' \dot{ A}_3-  A_3'\dot{ V}_1\right), \label{e:cpe4} \\
    0 & = 2\Delta V_0 - \rho  + 16\alpha B \int_0^1 \diff z \, z \,  A_3. \label{e:cpe5}
\end{align}
As in section \ref{sec:NESS}, we seek a steady state solution using an ansatz of the same functional form. The boundary data are now given by $\phi'(0) = M$, $A_3(0)=b$, $g_1(0)=P$, $g_2(0)=0$ and $\Delta V_0(t) = \mu$, where $\mu$ is interpreted as the chemical potential.

\begin{figure}[h]
\centering
\includegraphics[width=0.5\textwidth]{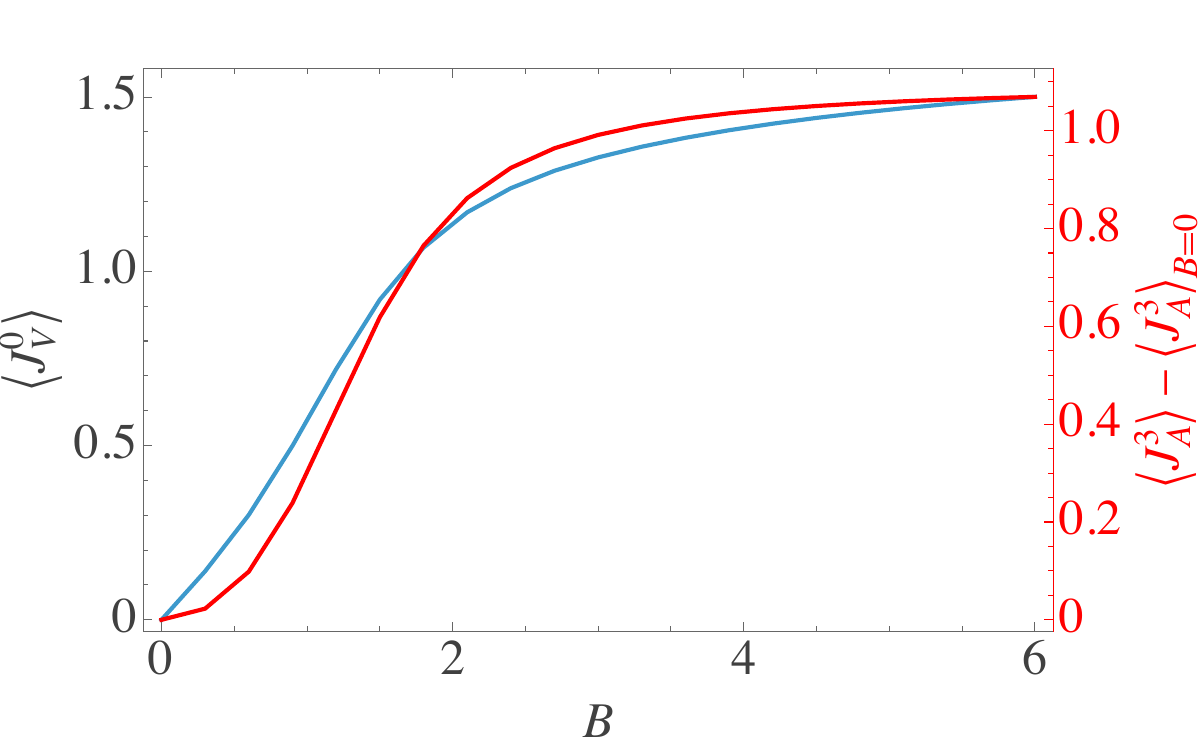}

\caption{Demonstration of the chiral pumping effect at vanishing chemical potential ($\mu=0$). The response of the charge density and axial current to an external magnetic field $\vec{B}=B \hat{e}_3$ is shown. 
\label{fig:CPE}}
\end{figure}

By repeatedly solving the system \eqref{e:cpe1}--\eqref{e:cpe4} for different values of $(\rho ,B)$ and then plugging the result into \eqref{e:cpe5} one numerically determines the functional dependence of the chemical potential $\mu = \mu(\rho ,B)$. Numerical inversion yields the dependence of the charge density $\rho  = \rho  (\mu,B)$. Fig.~\ref{fig:CPE} illustrates the resulting responses in the case of zero chemical potential ($\mu=0$). Notice that despite the vanishing chemical potential, there is a nonzero charge density accompanied by an axial current in the direction of the magnetic field, in agreement with the CPE. Moreover, the dependence of the charge density on the magnetic field is approximately linear at small $B$. This behavior can be related to the Dirac theory as follows. The axial gauge field admits an expansion in the  magnetic field of the form $A_3(z;B) = A_3^{(0)}(z)B^0 + A_3^{(1)}(z)B^1 + O(B^2)$. Plugging into \eqref{e:cpe5} with $\mu = 0$ and integrating by parts we obtain
\begin{align}
    \rho
    & = 8 \alpha B 
    \left[
    A_3^{(0)}(1) - \int_0^1 \diff z \, z^2 (A_3^{(0)})'(z) \right] + O(B^2).
\end{align}
Next, we recall that the Chern-Simons anomaly coefficient $\alpha$ is related to the charge in the Dirac theory by $8\alpha = \frac{q^2}{2\pi^2}$ \cite{landsteiner2016holographic}. Thus, identifying the effective chiral shift with $b_\textrm{eff} = A_3^{(0)}(1) - \int_0^1 \diff z \, z^2 (A_3^{(0)})'(z)$, then we obtain, at leading order in the magnetic field,
\begin{equation}
    \rho  = \frac{q^2 b_\textrm{eff} B}{2\pi^2},
\end{equation}
in agreement with \cite[Eq.~(15)]{ebihara2016chiral}. Interestingly, the NESS has generated an additional bulk contribution to the chiral shift compared to the static problem discussed in section \ref{sec:ahe}. Finally, we compare with the Dirac theory prediction that the axial current satisfies $\langle J_A^3 \rangle \propto \langle J_V^0 \rangle$ with some regulator-dependent coefficient (see \cite[Eq.~(17)]{ebihara2016chiral} and appendix \ref{app:CPE}). Plotting the excess axial current alongside the pumped charge in Fig.~\ref{fig:CPE}, we notice that although the axial current closely tracks the charge density, there appears to be strong violations to the proportionality $\langle J_A^3 \rangle \propto \langle J_V^0 \rangle$. This result is not surprising, however, given that the axial current is not protected by the anomaly and receives regulator-dependent contributions which can dominate at strong coupling.

\section{Discussion}\label{sec:discussion}

A nonequilibrium steady state (NESS) was constructed holographically for a Floquet-driven Weyl semi-metal in the probe limit model of \cite{landsteiner2016holographic}, where a rotating boundary electric field provides the periodic drive while the black-hole horizon supplies dissipation at fixed temperature. The resulting steady-state solutions exist over a wide range of driving parameters and provide a controlled, strongly coupled realization of Floquet dynamics in an open quantum system.

The dynamical stability of the NESS was characterized by the spectrum of linearized fluctuations. A critical curve in drive parameter space was identified by Floquet exponents crossing into the upper half of the complex plane. Beyond this curve the steady state destabilizes and the axial response develops superharmonic components, while the vector response remains commensurate with the drive. Numerical solution of the initial-boundary value problem corroborates this picture and shows that sufficiently strong driving in the unstable regime produces irregular time dependence indicative of chaotic dynamics. From the bulk perspective, the horizon prevents indefinite heating but does not preclude nonlinear instabilities.

The anomaly-induced response of the NESS to an external magnetic field was also investigated. In free Dirac theory the chiral pumping effect predicts that circular driving generates an effective axial background which, in the presence of a magnetic field, pumps vector charge into the sample and induces an axial current along the field direction \cite{ebihara2016chiral}. The holographic results exhibit the same qualitative structure: a magnetic field produces both a vector charge density and an axial current whose magnitude is controlled by the drive-generated axial background. This provides evidence that the chiral pumping mechanism persists at strong coupling.

Several extensions would be of interest. Incorporating an axial magnetic field, which is expected to source a longitudinal vector current $J_V^3$, would require leaving the spatially homogeneous ansatz, since the scalar couples directly to the axial gauge potential $A_\mu$ rather than only to its field strength and thus induces nontrivial $\vec{x}$-dependence. Going beyond the probe limit to include metric backreaction would clarify the fate of the strongly driven, apparently chaotic regime. In particular, including backreaction allows for studying quantum information theoretic quantities such as entanglement entropy. Since the two Weyl cones were originating from the same Dirac cone and have been separated by the electric field (in momentum space), it would be very interesting to understand the entanglement production in this process. Moreover, one could investigate whether the phase diagram of this nonequilibrium phase transition can be characterized through entanglement~\cite{Rangamani:2015sha,Biasi:2017kkn}. 

The authors of \cite{Gaschler:2025ayj} observed long lived oscillations in the current, even after the electric field is switched off. These resonances occur in strong magnetic fields and for sufficiently large values of the Chern-Simons coupling and were first observed in \cite{Ammon:2016fru} (see~\cite{Haack:2018ztx,Ghosh:2021naw,Grieninger:2023wuq,Grieninger:2023myf,Meiring:2023wwi,Waeber:2024ilt} for follow up works). It would be interesting to understand how the dynamics induced by these resonances affects the CPE and whether it can support an effective NESS (with some finite lifetime).

Finally, instead of breaking time-reversal symmetry by driving the system periodically with an electric field, Weyl semi-metals may also be induced by applying a strain \cite{Cortijo:2016wnf}. In a magnetic field the strain can induce an electric current via the chiral magnetic effect. It would be interesting to contrast the dynamics with our study.

\acknowledgments

We thank Martin Ammon and Markus Gardemann for useful discussions in the early stages of this work.
J.S.~thanks T\'{o}mas Andrade for sharing the code accompanying \cite{andrade2017holographic}. MB acknowledges the support of the Shanghai Municipal Science and Technology Major Project (Grant No.2019SHZDZX01). This work was supported in part
by the U.S. Department of Energy, Office of Science, Office of Nuclear Physics, Inqubator for Quantum Simulation (IQuS) under Award Number DOE
(NP) Award DE-SC0020970 (S.G.). This work was also supported by the U.S. Department of Energy, Office of Science, Office of Nuclear Physics, Grant No. DE-FG02-97ER-41014 (UW Nuclear Theory, S.G.). S.G. was supported in part by a Feodor Lynen Research fellowship of the Alexander von Humboldt foundation. This work was also supported, in part, by the Department of Physics and the College of Arts and Sciences at the University of Washington.

\appendix

\section{Free fermion computations}
In this section we depart from the convention of the main text by using the mostly minus metric convention.
\subsection{Current response}\label{app:superharmonic}
In this section we provide a free-fermion justification for the superharmonic axial response. Consider a Dirac field of mass $m$ in odd spatial dimension $d$ coupled to a nondynamical gauge field $V_a=V_a(t)$, which vanishes for $t\leq 0$. Consider the Schr\"{o}dinger picture with reference time chosen to be $t=0$. The Hamiltonian is given by
\begin{align}
    \hat{H} & = \hat{H}_0 + \hat{H}_1, \\
    \hat{H}_0 & = 
    \int_{\mathbb{R}^d} \diff^d \vec x \, \bar\psi(\vec x)\big({-\iu\vec\gamma\cdot\vec\nabla + m}\big)\psi(\vec x), \\
    \hat{H}_1(t)
    & =
    -V_a(t)\int_{\mathbb{R}^d} \diff^d \vec x \, \hat{J}_V^a(\vec{x}),
\end{align}
where the current operators are defined as follows,
\begin{align}
    \hat{J}_V^a(\vec{x}) & := \bar\psi(\vec x)\gamma^a \psi(\vec x), \\
    \hat{J}_A^a(\vec{x}) & := \bar\psi(\vec x)\gamma^a \gamma^5 \psi(\vec x),
\end{align}
and the Schr\"{o}dinger picture field operators satisfy
\begin{equation}
    \{ \psi_\alpha(\vec{x}), \psi_\beta^\dag(\vec{x}') \} = \delta_{\alpha\beta}\delta^d(\vec{x}-\vec{x}').
\end{equation}
Define the Heisenberg-picture fields by
\begin{equation}
    \psi_\textrm{H}(\vec{x},t) = \hat{U}^\dag(t,0)\psi(\vec{x})\hat{U}(t,0).
\end{equation}
Recall that
\begin{equation}
    \psi_\textrm{H}(\vec{x},t) = \hat{U}_\textrm{D}^\dag(t,0) \psi_\textrm{D}(\vec{x},t) \hat{U}_\textrm{D}(t,0),
\end{equation}
where
\begin{align}
    \hat{U}_\textrm{D}(t,0)
    & = \textrm{T} \exp \left[-\iu \int_0^t \diff u \, \hat{V}_\textrm{D}(u)\right],
\end{align}
and $\hat{V}_\textrm{D}(u)$ is obtained from $\hat{H}_1$ by replacing $\psi(\vec{x}) \to \psi_\textrm{D}(\vec{x},u)$. 

In the following, we compute the quantum expectation value of the current operators using time-dependent perturbation theory,
\begin{align}
    \mathcal{J}^a(t) := \langle 0| \hat{J}^a_\textrm{H}(\vec{x},t) |0\rangle,
\end{align}
where $\hat{J}^a = \hat{J}_{A,V}^a$. The Dyson series gives (henceforth omitting the subscripts on Dirac-picture fields),
\begin{align}
    \hat{J}^a_\textrm{H}(\vec{x},t)
    & =
    \hat{U}_\textrm{D}^\dag(t,0) 
    \hat{J}^a(\vec{x},t)
    \hat{U}_\textrm{D}(t,0), \\
    & = 
    \hat{J}^a(\vec{x},t)
    - 
    \iu \int_0^t \diff t' \, \big[\hat{J}^a(\vec{x},t), \hat{V}_\textrm{D}(t')\big]
    +
    \cdots, \\
    & = 
    \hat{J}^a(\vec{x},t)
    + 
    \iu \int_0^t \diff t' \, V_b(t')  \int_{\mathbb{R}^d} \diff^d \vec{x}' \, \big[\hat{J}^a(\vec{x},t), \hat{J}_V^b(\vec{x}',t')\big]
    +
    \cdots.
\end{align}
Thus,
\begin{align}
    \mathcal{J}^a(t)
    & = 
    \langle 0| \hat{J}^a(\vec{x},t) |0\rangle
    + 
    \iu \int_0^t \diff t' \, V_b(t')  \int_{\mathbb{R}^d} \diff^d \vec{x}' \, \langle 0| \big[\hat{J}^a(\vec{x},t), \hat{J}_V^b(\vec{x}',t')\big] |0\rangle
    +
    \cdots.
\end{align}
The first term vanishes by Poincar\'{e} invariance. Now specialize to $d=3$. If $\hat{J}^a=\hat{J}_A^a$, then the second term vanishes by charge conjugation invariance. Thus, the first nontrivial term in $\mathcal{J}_V^a$ and $\mathcal{J}_A^a$ begin at linear and quadratic order in $V_a$, respectively. Suppose that $V_a(t) \propto (0,\cos\omega_\textrm{D},\sin\omega_\textrm{D},0)$. Then it follows that $\mathcal{J}_V^a$ and $\mathcal{J}_A^a$ are dominated by frequencies $\omega=\omega_\textrm{D}$ and $\omega\in \{ 0,2\omega_\textrm{D}\}$, respectively. Note that the existence of the DC term in $\mathcal{J}_A^a$ allows the formation of a nonzero time average $\langle \mathcal{J}_A^a \rangle \neq 0$.

\subsection{Chiral pumping effect from first quantization}\label{app:CPE}
Consider the Weyl representation of the Clifford algebra,
\begin{align}
    \sigma^a & = (1, \vec\sigma), & \bar{\sigma}^a & = (1, -\vec\sigma), \\
    \gamma^a & =
    \begin{bmatrix}
        0 & \sigma^a \\
        \bar{\sigma}^a & 0
    \end{bmatrix},
    &
    \gamma^5 
    & = \iu \gamma^0 \gamma^1 \gamma^2 \gamma^3
    =
    \begin{bmatrix}
    -1 & 0 \\
    0 & 1
    \end{bmatrix}.
\end{align}
If we define spinors $\phi_\pm$ by
\begin{equation}
    \psi
    =
    \begin{bmatrix}
    \phi_- \\
    \phi_+
    \end{bmatrix},
\end{equation}
then the massless Dirac equation with external vector/axial gauge fields becomes
\begin{align}
    \sigma^a \big[\iu \partial_a + q\big(V_a + A_a\big)\big] \phi_+
    & = 0, \\
    \bar\sigma^a \big[\iu \partial_a + q\big(V_a - A_a\big)\big] \phi_-
    & = 0,
\end{align}
where we have chosen $q_5=q$. Assuming $V_0=A_0=0$ and passing to Hamiltonian form gives
\begin{align}
    \iu \partial_t \phi_+ &  = - \iu \vec{\sigma} \cdot \vec{\nabla} \phi_+ - \vec{\sigma}\cdot q\big(\vec{V} + \vec{A}\big)\phi_+, \\
    \iu \partial_t \phi_- &  = \iu \vec{\sigma} \cdot \vec{\nabla} \phi_- + \vec{\sigma}\cdot q\big(\vec{V}-\vec{A}\big)\phi_-.
\end{align}
Now consider the following backgrounds (with $B>0$),
\begin{align}
    \vec{V} & = (0, Bx, 0), \\
    \vec{A} & = \big(0, 0, \tfrac{b_\textrm{eff}}{q}\big).
\end{align}
Consider the following ansatz,
\begin{equation}
    \phi = e^{-\iu (\omega t  - k_y y - k_z z)}
    \begin{bmatrix}
        u(x) \\
        d(x)
    \end{bmatrix}.
\end{equation}
If we assume that $|q|B\gg \Lambda^2$ then the physics is described by the lowest Landau level (LLL). Now suppose $q >0$. It can be shown that the LLL solution is
\begin{align}
    u_\pm(x)
    & = e^{-\frac{qB}{2}(x-x_0)^2}, \\
    d_\pm(x)
    & = 0, \\
    x_0
    & = \frac{k_y}{qB},
\end{align}
with dispersion relation
\begin{equation}
    \omega_\pm(k_y,k_z;b_\textrm{eff}) = \pm k_z - b_\textrm{eff}.
\end{equation}
Now compactify the $x$-$y$ plane on a 2-torus of size $L_xL_y$. Then the requirement that $x_0 \in (0,L_x]$ imposes the wave number condition $k_y \in K_y$, where
\begin{align}
    K_y & = \left\{\frac{2\pi n_y}{L_y} : n_y = 1,\ldots, N_B\right\}, \\
    N_B & = \frac{qB}{2\pi}L_xL_y.
\end{align}
The total charge density in the ground state is given by counting the negative-energy states. In order to facilitate counting, we compactify the $z$-direction on a circle of circumference $L_z$ so that the total volume is given by $V=L_xL_yL_z$. Then the charge density is given by the infinite expression
\begin{equation}
    J_V^0(b_\textrm{eff}) = \frac{q}{V}\sum_{k_y\in K_y}\sum_{k_z \in K_z}
    \Big[
    \theta\big({-\omega_+(k_y,k_z;b_\textrm{eff})}\big)
    +
    \theta\big({-\omega_-(k_y,k_z;b_\textrm{eff})}\big)
    \Big],
\end{equation}
where $K_z = \frac{2\pi}{L_z}\mathbb{Z}$.
The physical charge density $J_V^0$ is obtained by subtracting the infinite Dirac sea contribution with $b_\textrm{eff}=0$. In the limit $L_z \to \infty$ we may replace $\sum_{k_z \in K_z} \to \frac{L_z}{2\pi}\int_{-\infty}^\infty \diff k_z$ and thus
\begin{align}
    J_V^0
    & = J_V^0(b_\textrm{eff})- J_V^0(0), \\
    & = \frac{q}{V} \left[N_B \left(\frac{L_z}{2\pi}\int_{-b_\textrm{eff}}^{b_\textrm{eff}} \diff k_z\right)\right], \\
    & = \frac{q^2 b_\textrm{eff} B}{2\pi^2}.
\end{align}
Next, define projectors
\begin{equation}
    P_\pm = \frac{1}{2}(I \pm \sigma^3 \otimes \sigma^3),
\end{equation}
which satisfy
\begin{equation}
    P_\pm(\gamma^0 \gamma^3 \gamma^5)P_\pm = \pm P_\pm.
\end{equation}
Recall that
\begin{align}
    J_V^0 & = q \bar{\psi} \gamma^0 \psi = q \psi^\dag \psi, \\
    J_A^3
    & = q \psi^\dag \gamma^0 \gamma^3 \gamma^5 \psi.
\end{align}
Since $\psi = P_+ \psi$, we obtain
\begin{equation}
    J_A^3 = J_V^0.
\end{equation}
Now suppose $q < 0$. Then the LLL is given by
\begin{align}
    u_\pm(x) & = 0, \\
    d_\pm(x) & = e^{\frac{qB}{2}(x-x_0)^2}, \\
    x_0 & = \frac{k_y}{qB},
\end{align}
with dispersion relation
\begin{equation}
    \omega_\pm(k_y,k_z;b_\textrm{eff}) = \mp k_z + b_\textrm{eff}.
\end{equation}
The charge density is again given by
\begin{align}
    J_V^0
    & = \frac{q^2 b_\textrm{eff} B}{2\pi^2},
\end{align}
but now $\psi = P_- \psi$ so we obtain
\begin{equation}
    J_A^3 = -J_V^0.
\end{equation}
Therefore in general,
\begin{equation}
    \boxed{J_A^3 = \sgn(q) J_V^0}
\end{equation}

\section{Holographic renormalization}\label{app:renormalization}
Recall that,
\begin{align}
    (\delta S)_\textrm{os}
    & =
    -
    \int \diff^5 x \, \partial_\nu \left[ \sqrt{-g}\left( H^{\nu\mu} + 4\alpha \epsilon^{\mu\nu\rho\lambda\sigma} A_\rho H_{\lambda\sigma} \right)\delta V_\mu\right] + \cdots,
\end{align}
which implies
\begin{align}
    \left.\frac{\delta S}{\delta V_0}\right|_\textrm{os}
    & =
    \left.\left[8 \alpha B  A_3 -\frac{1}{\epsilon}  V_0'\right]\right|_{z=\epsilon}, \\
    & = \rho .
\end{align}
The covariant counter-terms are given by
\begin{align}
    S_{\rm ct}[\gamma] & = \int \diff^4 x \sqrt{-\gamma} \left\{ 
    \left[\gamma^{ab}\phi^\ast D_a D_b \phi-\frac{1}{4} F^2-\frac{1}{4}H^2\right] \log\epsilon
    -
    \phi^\ast \phi
    \right\}.
\end{align}
The corresponding functional derivative is given by
\begin{equation}
    \frac{\delta S_\textrm{ct}}{\delta V_a(\epsilon)}[\gamma]
    = \partial_b \big( \sqrt{-\gamma} \gamma^{bc}\gamma^{ad} H_{cd} \big) \log\epsilon.
\end{equation}
The background value of the $\gamma$ is given by
\begin{align}
    (\gamma_\textrm{bg})_{ab}
    & = 
    \begin{bmatrix}
    -f(\epsilon) & 0 \\
    0 & I_3
\end{bmatrix}.
\end{align}
In the case of Schwarzschild AdS$_5$ for which $f(z)=1-z^4$, we have the following important simplification,
\begin{align}
    (\gamma_\textrm{bg})_{ab}
    & = \frac{1}{\epsilon^2}\eta_{ab} + O(\epsilon^2).
\end{align}
Thus, evaluating on the background metric,
\begin{align}
    \frac{\delta S_\textrm{ct}}{\delta V_a(\epsilon)}[\gamma_\textrm{bg}]
    & = \big(
    \square V^a - \partial^a \partial_b V^b
    \big)\log\epsilon + \cdots, \\
    & = \big(
    \square V^a - \partial^a \dot{V}^0
    \big)\log\epsilon + \cdots.
\end{align}
Thus,
\begin{align}
    \frac{\delta S_\textrm{ct}}{\delta V_0(\epsilon)}[\gamma_\textrm{bg}]
    & = -\big(
    \ddot{V}^0 + \partial^0 \dot{V}^0
    \big)\log\epsilon + \cdots, \\
    & = 0+\cdots, \\
    \frac{\delta S_\textrm{ct}}{\delta V_j(\epsilon)}[\gamma_\textrm{bg}]
    & = - \ddot{V}^j \log\epsilon + \cdots.
\end{align}
It follows that 
\begin{equation}
    \langle J_V^0 \rangle_\textrm{ren} = \rho .
\end{equation}
Next, recall that,
\begin{align}
    (\delta S)_\textrm{os}
    & =
    -
    \int \diff^5 x \, \partial_\nu \left[\sqrt{-g}\left(F^{\nu\mu} + \frac{4}{3}\alpha \epsilon^{\mu\nu\rho\lambda\sigma} A_\rho F_{\lambda\sigma} \right)\delta A_\mu\right] + \cdots,
\end{align}
and
\begin{equation}
    \frac{\delta S_\textrm{ct}}{\delta A_a(\epsilon)}[\gamma]
    = 
    \Big[
    \partial_b \big( \sqrt{-\gamma} \gamma^{bc}\gamma^{ad} F_{cd} \big) 
    +
    \iu q \sqrt{-\gamma} \gamma^{ab}\big(\phi (D_b\phi)^\ast - \phi^\ast D_b\phi\big)
    \Big]
    \log\epsilon.
\end{equation}
It follows that
\begin{equation}
    \langle J_A^3 \rangle_\textrm{ren} = M^2 b q^2 + 2 a(t).
\end{equation}

\section{Initial-boundary value problem formulation}\label{app:IBVP}
In this section we describe some details of the initial-boundary value problem with $B=0$ (and thus $V_0=0$).
Evaluating the equations of motion at $z=1$ produces a complicated IR boundary condition. It is convenient to instead work with the coordinate $\rho$ defined by $z=1-\rho^2$. Then evaluating the scalar equation of motion at $z=1$ ($\rho = 0$) we obtain
\begin{equation}
    \partial_t\partial_\rho\varphi(t,0) + \partial_\rho \varphi(t,0) = 0,
\end{equation}
which is solved by
\begin{equation}
    \partial_\rho \varphi(t,0) = c e^{-t},
\end{equation}
where $c$ is an integration constant. The only physically acceptable integration constant is $c=0$ and thus we obtain the simple Neumann condition
\begin{equation}
    \partial_\rho \varphi(t,0)=0.
\end{equation}

Next we consider the gauge fields expressed in terms of the $\rho$ coordinate, which we denote by $a_3,v_1,v_2$. Evaluating the $a_3$, $v_1$ and $v_2$ equations of motion at $\rho = 0$ gives, respectively
\begin{align}
    \partial_t\partial_\rho a_3(t,0) + \partial_\rho a_3(t,0)
    & =
    4\alpha\big[\partial_t v_1(t,0)\partial_\rho v_2(t,0) - \partial_\rho v_1(t,0)\partial_t v_2(t,0)\big], \\
    \partial_t\partial_\rho v_1(t,0) + \partial_\rho v_1(t,0)
    & = 
    4\alpha\big[\partial_t v_2(t,0)\partial_\rho a_3(t,0) - \partial_\rho v_2(t,0)\partial_t a_3(t,0)\big],
    \\
    \partial_t\partial_\rho v_2(t,0) + \partial_\rho v_2(t,0)
    & =
    4\alpha\big[\partial_t a_3(t,0)\partial_\rho v_1(t,0) - \partial_\rho a_3(t,0)\partial_t v_1(t,0)\big].
\end{align}
If we define 
\begin{align}
    \mathbf{w}(t)
    & = 
    \begin{bmatrix}
    \partial_\rho a_3(t,0) \\
    \partial_\rho v_1(t,0) \\
    \partial_\rho v_2(t,0)
    \end{bmatrix},
    &
    \mathbf{S}(t)
    & = 
    \begin{bmatrix}
    \partial_t a_3(t,0) \\
    \partial_t v_1(t,0) \\
    \partial_t v_2(t,0)
    \end{bmatrix},
\end{align}
then the IR boundary conditions become
\begin{equation}
    \dot{\mathbf{w}}(t) = 4 \alpha \mathbf{S}(t) \times \mathbf{w}(t) - \mathbf{w}(t).
\end{equation}
Taking the dot product with $\mathbf{w}(t)$ we obtain
\begin{align}
    \mathbf{w}(t) \cdot \dot{\mathbf{w}}(t) & = - \Vert \mathbf{w}(t) \Vert^2 \implies \\
    \frac{1}{2} \partial_t \Vert \mathbf{w}(t) \Vert^2 & = - \Vert \mathbf{w}(t) \Vert^2 \implies \\
    \Vert \mathbf{w}(t) \Vert^2 & = c e^{-2t}.
\end{align}
The only physically acceptable integration constant is $c=0$ and thus $\Vert \mathbf{w}(t) \Vert = 0 \implies \mathbf{w}(t)=0$. Therefore, the complete set of IR conditions is of Neumann form,
\begin{align}
    \partial_\rho \varphi(t,0) & = 0, & \partial_\rho a_3(t,0) & = 0, \\
    \partial_\rho v_1(t,0) & = 0, & \partial_\rho v_2(t,0) & = 0.
\end{align}

\bibliographystyle{JHEP}
\bibliography{biblio}

\end{document}